\documentstyle[preprint,aps]{revtex}
\tightenlines
\begin{document}
\pagestyle{empty}
\draft
\centerline{hep-ph/9303226 \hfill CERN--TH.6814/93}

\begin{center}
{\Large
CP--odd Correlation in the Decay of Neutral Higgs Boson  \\
into $ZZ$, $W^+W^-$, or $t{\bar t}$
}
\end{center}
\vskip 2cm
\centerline{
Darwin Chang$^{(1)}$,
Wai--Yee Keung$^{(2,3)}$ and
Ivan Phillips$^{(1)}$}
\vskip 1cm
\begin{center}
$^{(1)}$Department of Physics and Astronomy,\\
Northwestern University, Evanston, IL 60208 \\
$^{(2)}$Theory Group, CERN
CH-1211, Geneva 23, Switzerland\\
$^{(3)}$Physics Department, University of Illinois at Chicago, IL 60607--7059\\
\end{center}
\vskip -1cm
\begin{abstract}
We investigate the possibility of detecting CP--odd angular correlations
in the various decay modes of the neutral Higgs boson including the modes
of  a $ZZ$ pair, a $W^+W^-$  pair, or a heavy quark pair.
It is a natural way to probe the CP character of
the Higgs boson once it is identified.
Final state interactions ({\it i.e.} the absorptive decay amplitude)
is not required in such correlations.
As an illustrative example we take the fundamental source of the
CP nonconservation to be in the Yukawa couplings of the Higgs boson
to the heavy fermions.
A similar correlation in the process $e^+e^- \rightarrow l^+ l^- H$ is
also proposed.
Our analysis of these correlations will be useful for experiments
in future colliders such as LEP II, SSC, LHC or NLC.
\end{abstract}
\vskip 1cm

\centerline{
Published in Phys. Rev. D{\bf 48}, 3225 (1993).}
\vfill
\narrowtext
\pagestyle{plain}
\section{Introduction}
The Higgs boson sector remains the most mysterious part of theories of
electroweak unification.
To probe this illusive sector one is well advised
to keep an open mind.
Most analyses of the Higgs sector, in the Standard
Model or beyond, ignore potential CP violation.  However, in many these
models, even the lightest neutral Higgs boson can have interesting CP
violating phenomena.
In many models of CP violation, including the Standard
Model, CP violation is a consequence of simultaneous existence of
many coupling constants.
Such requirement results in the presence of
many coupling constants and/or loop
suppression factors in the observables.
The neutral Higgs sector is unique
in the sense that a single Higgs boson coupling to a massive fermion is
enough to manifest CP violation as long as the Yukawa coupling contains
both scalar and pseudoscalar components.  Therefore, in many ways the CP
violating aspect may be the most interesting part of the Higgs boson
physics beyond the the Standard Model once a neutral Higgs is identified.
 
In a previous paper, we investigated the signatures of
various CP--odd asymmetries
in different polarized decay modes the neutral Higgs boson\cite{ref:ck1}.
Among the interesting modes are the CP asymmetries in the
event rate differences $N(Q_L {\bar Q}_L)-N(Q_R {\bar Q}_R)$,
$N(W_L^+ W_L^-) - N(W_R^+ W_R^-)$ or $N(Z_L Z_L) - N(Z_R Z_R)$.
As expected, the CP asymmetries will manifest themselves only after the
final state interactions are taken into account.
For the heavy quark pair, $Q\bar Q$, mode or the $W^+W^-$ pair mode, the
relative energy of the final state charged leptons can be used as the
polarization analyzers of the heavy quarks or the $W$'s
\cite{ref:kly,ref:Peskin}.
However for the
case of the $ZZ$ mode, the $Z$ coupling to charged leptons is mainly
axial-vectorial.  Therefore the efficiency of using its leptonic
energy spectrum as the polarization analyzer is
suppressed by a factor of $c_V/c_A = -1 + 4\sin^2\theta_W \sim -0.08$.
 
In this paper we consider a different kind of signal of CP violation:
the CP--odd correlations of final state momenta.  Note that this is
different from the more conventional T-odd signals
which can be imitated by
CP conserving final state interactions.  No final state interaction is
required to obtain the signals.   Therefore,
unlike the case of the CP asymmetry
considered in Ref. \cite{ref:ck1},
it is not necessary to cross a heavy fermion
threshold to obtain a large signal in the $W^+W^-$ or the $ZZ$ channel.
For earlier discussions of CP--odd effects in different processes,
see Refs.\cite{ref:early,ref:eeWW,ref:eett,ref:CPjet,ref:ckp}.
 
\section{$H\to ZZ$}
Let us start with the $ZZ$ mode whose CP--odd asymmetry
does not translate
into large lepton energy asymmetry.
After the $Z$'s decay there are three modes of final state: (1)
$l^- l^+ l'^- l'^+$, (2) $l^- l^+ q \bar{q}$,
(3) $q \bar{q} q' \bar{q'}$.
The purely leptonic modes are of course the easy ones
in which to get a CP--odd
correlation provided that we collect enough events.
In the bigger samples of the semileptonic modes and the hadronic ones,
it is difficult to tell the charge of the leading quark when it
hadronizes into a jet.
However, the most intriguing part of our result is that it is
not necessary to identify charge to detect the CP--odd signal,
an argument
we will elaborate on later.
This observation will be even more crucial in trying to decode CP--odd
signals in the $W^+ W^-$ or $t \bar{t}$ modes of the Higgs decay.
Another important point is that in modes (1) and (3),
the lepton pairs or the
quark pairs do not have to be distinct.  As we will show later,
the identical particle effect does not wash out the main CP--odd signal.
We shall discuss these modes in order.
 
Consider $H \rightarrow ZZ \rightarrow l^- l^+ l'^- l'^+$.
If $H$ is scalar, the two $Z$
polarizations will be parallel; if $H$ is pseudoscalar, the two $Z$
polarizations will be perpendicular.  The correlation between the
polarizations of the two $Z$ bosons translates into
the correlation between the
two planes of the $l^+l^-$ pairs.
In fact this observation was used a
long time ago to measure the parity of the $\pi^0$
through its decay into two
photons\cite{ref:wolf}.
When the Higgs coupling is neither scalar
nor pseudoscalar, one has a source of
CP violation.  We shall explore this case.
 
The effective interaction for the $ZZH$ vertex can be written as
\begin{equation}
{\cal L}_{HZZ}= \case1/2 g_H H
[ B Z^\nu Z_\nu + \case1/4 D M_H^{-2} \epsilon_{\mu\nu\rho\sigma}
                        Z^{\mu\nu} Z^{\rho\sigma}] \; .
\end{equation}
where $g_H = (g_2 M_Z / \cos\theta_W)$,
$Z^{\mu\nu}$ is the field strength of $Z$ boson and $B$ and $D$ are
dimensionless form factors.  For the Higgs boson in the Standard Model,
$D=0, B=1$.
The $D$ term is CP--odd while the $B$ term is CP--even.
Simultaneous presence of $D$ and $B$ is CP violating.
We have ignored higher dimensional CP-even operators in the above
Lagrangian and a possible CP violating
$H Z \gamma$ vertex which could also contribute to four fermions final
states.
Their effect should be negligible at the pole of the final state Z
boson.
In momentum space, these effective vertices give rise
to the covariant amplitude:
\begin{equation}
iM_{H\rightarrow Z(P,\eta),Z(P',\eta')}
= ig_H[B \eta\cdot\eta' +
(D/M_H^2)\epsilon(\eta,\eta',P,P')] \;.
\end{equation}
Here  $\eta, \eta'$ are the polarizations and $P, P'$ are the
momenta of the two $Z$ bosons. We define
$\epsilon(\eta,\eta',P,P') \equiv
\epsilon_{\mu\nu\alpha\beta}\eta^\mu \eta'^\nu P^\alpha P'^\beta$,
with the convention $\epsilon_{0123}=1$.
The helicity amplitudes in $ H \rightarrow ZZ$ are represented as
\begin{eqnarray}
M_{+,+} =  g_H(B - \case{i}/2 D\beta), \
M_{-,-} =  g_H(B + \case{i}/2 D\beta), \
M_{0,0} = -g_HB {1+\beta^2 \over 1-\beta^2}    \;,
\label{eq:MZZ}
\end{eqnarray}
in the  $H$ rest frame, with $\beta^2=1-4M_Z^2/M_H^2$.
Here the subscripts $(+,-,0)$ denote respectively the helicities
$(R,L,\parallel)$ of the $Z$ bosons.
Conservation of angular momentum implies only the above three helicity
configurations for this decay process.
Under CP transformation,
$M_{+,+} \rightarrow M_{-,-}$, and $M_{00} \rightarrow M_{00}$.
Thus, $D$ is CP--odd, however Re~$D$ is CP$\hat{T}$--even,
while Im~$D$ is CP$\hat{T}$--odd.
The detail definition of CP$\hat{T}$ can be found in Ref.\cite{ref:ckp}.  
It is roughly the CPT symmetry without reversing the initial and the final 
states, that is, the symmetry reverses only the kinematic variables of 
the states according to their CPT property.  The
effect of Im~$D$ will give an asymmetry in the production of polarized
states,
\begin{equation}
N(Z_LZ_L)-N(Z_RZ_R)=(-\beta \hbox{Im }D/B)[N(Z_LZ_L)+N(Z_RZ_R)] \;.
\end{equation}
Its consequence in the energy asymmetry of the final leptons
has been discussed in Ref.\cite{ref:ck1}.
In this paper, we focus our attention only on the effect of the
real part of $D$ through the angular asymmetry of the final decay
products.
In the decay process $H \rightarrow ZZ \rightarrow
l^-(\vec{k}_-) l^+(\vec{k}_+) l'^-(\vec{p}_-) l'^+(\vec{p}_+)$,
the angular correlation of the final 4 fermions  is encoded in the
$3\times3$ density matrix
\begin{equation}
\rho^{\lambda}_{\lambda'} =
M_{\lambda, \lambda}\;M_{\lambda', \lambda'}^*
\quad \hbox{(no dummy summation),}
\label{eq:rho}
\end{equation}
which is folded with the $Z$ decay amplitudes
to produce the event distribution.
The CP--odd, CP$\hat{T}$--even combinations are $\rho^+_- - \rho^-_+$
and $\rho^0_+ - \rho^0_- + \rho^-_0 - \rho^+_0$ while the ones that are
CP-- and CP$\hat{T}$--odd are $\rho^+_+ - \rho^-_-$ and
$\rho^0_+ - \rho^0_- - \rho^-_0 + \rho^+_0$.
Only the first two are relevant
to our discussion below.
 
Similar notation will be used for the modes of $W^+$--$W^-$.
%
\section{CP violating observables}
We will show observables which are related to the
CP--odd correlations among the momenta,
$\vec{k}_-, \vec{k}_+, \vec{p}_-$ and $\vec{p}_+$.
To simplify our discussion, we start with the purely leptonic case
and we arbitrarily label the lepton pair
from one of the $Z$ bosons with primes.
The process $Z\rightarrow l\bar l$ can be parametrized by the vertex
$$i\,e\,\bar{u}(l)\gamma_\mu(c_L L + c_R R) v(\bar{l})\;.$$
Naively one may simply construct the CP--odd correlation
${O}_{odd} = \vec{p}_-\times \vec{p}_+\cdot \vec{k}_-$
in the $H$ rest frame.  It can also be written in a Lorentz invariant
form
$${O}_{odd} = - M_H^{-1}\epsilon(p_-,p_+,k_-,k_+) \;. $$
However, as we will show later,
the expectation value of this observable
$\langle \vec{p}_-\times \vec{p}_+\cdot \vec{k}_-\rangle $
is proportional to $c_V c_A$ where $c_V = \case1/2 (c_L + c_R)$ and
$c_A = \case1/2 (-c_L + c_R)$ are the vector and axial vector
couplings of the $Z$ boson.
Since the vector coupling of the $Z$ boson to the
charged leptons in the Standard Model is relatively small,
this observable turns out to be rather unimportant.
It can be understood as the consequence of
an approximate symmetry when the vector coupling
(or the axial vector coupling) is ignored completely,
so that there is no distinction between $l^+$  and $l^-$
as far as the $Z$ boson is concerned.
Therefore the differential decay rate is symmetric under two
separate partial charge conjugation symmetries,
$\hat{C}_1$ and $\hat{C}_2$.
Symmetry $\hat{C}_1$ interchanges $l^+$ and $l^-$ while leaving
$l'^+$, $l'^-$ unchanged; while the symmetry $\hat{C}_2$ interchanges
$l'^+$ and $l'^-$ and leaving $l^+$, $l^-$ unchanged instead.
The usual charge
conjugation operation $C$ is the product of the the two $\hat{C}$'s.
It is easy to check that the correlation $O_{odd}$ is odd under
either $\hat{C}_1$ or $\hat{C}_2$.
Therefore it has nonvanishing expectation value only when
both $c_V$ and $c_A$ couplings are nonzero.
It is certainly more desirable to use an
observable which is CP--odd but even under each $\hat{C}_1$ or
$\hat{C}_2$. It is not too difficult to construct a quantity.
A possibility is
\begin{equation}
  O'_{odd} = (\vec{p}_-\times \vec{p}_+\cdot \vec{k}_-)
[(\vec{p}_-\times \vec{p}_+)\cdot(\vec{k}_-\times \vec{k}_+)] .
\label{eq:op}
\end{equation}
 
It is easier to understand these angular correlations
in the geometry of the reaction.
A typical reaction is shown in Fig.~1.
Under CP transformation, $(E_{p_-}, E_{k_-}, \phi)$ is transformed into
$(E_{p_+}, E_{k_+}, -\phi)$.  If one defines a plane using
the cross product of one of the lepton pairs, say
$\vec{p}_-\times \vec{p}_+$, as the direction of the normal vector,
the configurations with the momentum of the other lepton ($k_-$)
coming out of the plane is the CP conjugate of the configurations
in which $k_-$ is going into the plane.
Therefore the asymmetry, $\langle sign(O_{odd})\rangle $,
can be interpreted as a kind of up-down asymmetry.
More explicitly,
define the polar and azimuthal angles for the two leptons
in their respective $Z$ boson rest frames to be
$\theta$,  $\phi$  for $l(\vec k_-)$,  and
$\theta'$, $\phi'$ for $l'(\vec p_-)$.
Here all the polar angles are defined
relative to the same $\hat{z}$ axis,
say the $\vec k_- + \vec k_+$ axis.
The distribution depends on the relative azimuthal angle, therefore
we can set azimuthal angle of $l'$ to $\phi'=0$ as shown in the Fig.~1.
Then one can label the event configuration by the angles
$(\cos\theta, \cos\theta', \phi)$.
The energies $E_{k_-}$ and $E_{p_-}$ in the Higgs boson rest frame
for the leptons $l$ and $l'$ respectively are determined by their polar
angles $\theta$ and $\theta'$.
Under $\hat{C}_1$, $\hat{C}_2$, parity and CP, the configuration
transforms as follows,
\begin{eqnarray}
\hat{C}_1&:&  { }(\cos\theta, \cos\theta', \phi) \rightarrow
              { }(\cos(\pi-\theta), \cos\theta', \phi+\pi) \nonumber \\
\hat{C}_2&:&  { }(\cos\theta, \cos\theta', \phi) \rightarrow
              { }(\cos\theta, \cos(\pi-\theta'), \phi+\pi) \nonumber \\
\hbox{P} &:&  { }(\cos\theta, \cos\theta', \phi) \rightarrow
              { }(\cos\theta, \cos\theta', -\phi) \nonumber \\
\hbox{CP}&:&  { }(\cos\theta, \cos\theta', \phi) \rightarrow
              { }(\cos(\pi-\theta), \cos(\pi-\theta'), -\phi).
\end{eqnarray}
To understand observables such as $\langle O'_{odd}\rangle $,
one can divide the azimuthal angle $\phi$ into four quadrants,
I, II, III, IV.
Since $\hat{C}_1$ transforms (I, II) into (III, IV),
it is clear that the up--down
asymmetry $\langle sign(O_{odd})\rangle $ which corresponds
to the angular integration of
(I + II) $-$ (III + IV) is odd under $\hat{C}_1$ or $\hat{C}_2$.
On the other hand, an observable similar to
$\langle sign(O'_{odd})\rangle $ can be constructed
by the alternative angular integration (I+III)-(II+IV).
\section{Angular dependence}
To calculate these asymmetries, we need the differential distribution:
\begin{eqnarray}
n(E_{p_-}, E_{k_-}, \phi) &\equiv&
{dN \over dE_{p_-} dE_{k_-} d\phi}             \nonumber\\
&=& {\cal N} \sum_{h,h'} \sum_{\lambda, \lambda'}
\rho^{\lambda}_{\lambda'}
f^h_{\lambda}(\theta,\phi)f^{h'}_{-\lambda}(\theta',0)
f^{h*}_{\lambda'}(\theta,\phi)f^{h'*}_{-\lambda'}(\theta',0)
c_h^2 c_{h'}^2 \;.
\label{eq:join}
\end{eqnarray}
We denote the helicity amplitude $f^h_{\lambda}(\theta,\phi)$ describing
the process $Z(\lambda) \rightarrow l(h) \bar l (-h)$ in the $Z$ rest
frame, where the spin projection of the $Z$ boson
along the $\hat z$ axis is
$\lambda\hbar$ and the helicity of $l$ is specified by $h=+(R)$
or $-(L)$.
One can relate the helicity amplitudes with the
spin--1 rotation matrix elements,
$f^h_\lambda \sim d^{\ 1}_{\lambda,h} e^{i\lambda\phi}$,
\begin{equation}
f^h_\pm (\theta,\phi)
    =(1\pm h\cos\theta)e^{\pm i \phi}/2 \;,   \quad
f^{h}_0=h\sin\theta/\sqrt{2}   \;,\quad   (h=\pm1)      \;.
\label{eq:damp}
\end{equation}
${\cal N}$ is the normalization to be specified later.
One can see from Eq.(\ref{eq:join},\ref{eq:damp}) that in order to get
the complex phase in angular distribution to expose the
CP--odd, CP$\hat{T}$--even effect, only the combinations
$\rho^+_- - \rho^-_+$ and $\rho^0_+ - \rho^0_- + \rho^-_0 - \rho^+_0$
contribute.  That is consistent with our general discussion after
Eq.(\ref{eq:rho}).
Note that in the $H \to ZZ \to l\bar l l\bar l$ mode of identical final
lepton pairs, the identical--particle symmetry implies
$n(E_{p_-}, E_{k_-}, \phi)=n(E_{k_-}, E_{p_-}, \phi)$.
Therefore it does not pose any restriction on the azimuthal,
$\phi$, angular distribution at all.
The overall distribution can be divided into two parts,
namely, the CP--even
piece $n_0$ and the CP--odd piece $\Delta n$,
\begin{equation}
n(E_{p_-}, E_{k_-}, \phi) = n_0     (E_{p_-}, E_{k_-}, \phi)
                           +\Delta n(E_{p_-}, E_{k_-}, \phi)
\label{eq:defevenodd}
\end{equation}
Using the form factors $D$, $B$ and the couplings $c_L$, $c_R$, we obtain
\begin{eqnarray}
\Delta n(E_{p_-}, E_{k_-}, \phi)
&=&{\cal N} D M_H^{-2} \epsilon(p_-,p_+,k_-,k_+) \nonumber\\
&\times&
[-(c_L^4 + c_R^4)(p_-.k_- + p_+.k_+) + 2c_L^2c_R^2(p_-.k_+ + p_+.k_-) ]
\;,                                    \nonumber\\
n_0(E_{p_-}, E_{k_-}, \phi)&=&  {\cal N}
B[(c_L^4 + c_R^4)p_-.k_-\, p_+.k_+ + 2c_L^2c_R^2p_-.k_+\, p_+.k_-] \;.
\label{eq:master}
\end{eqnarray}
From now on, except for the Section IX,
it is understood that we have dropped  the Re prefix for
the form factor $D$. We have only kept the linear piece in $D$, as a
result of perturbation.
Kinematically, we can express the $\epsilon$ symbol and
various scalar products in terms of angles, $\theta,\phi,\theta'$,
\begin{eqnarray}
\epsilon(p_-,p_+,k_-,k_+) =
- \case1/8 M_H^2 M_Z^2\beta\sin\theta' \sin\theta \sin\phi ,
\quad\quad\quad\quad\ \quad\quad\  \nonumber\\
p_-\cdot k_- = {M_H^2\over 16} \left[
(1+\beta^2)(1-\cos\theta \cos\theta')
+ 2\beta(\cos\theta - \cos\theta')\right]
- {M_Z^2\over 4} \sin\theta \sin\theta' \cos\phi ,\nonumber\\
p_+\cdot k_+ = {M_H^2\over 16} \left[
(1+\beta^2)(1-\cos\theta \cos\theta')
- 2\beta(\cos\theta - \cos\theta')\right]
- {M_Z^2\over 4} \sin\theta \sin\theta' \cos\phi ,\nonumber\\
p_-\cdot k_+ = {M_H^2\over 16} \left[
(1+\beta^2)(1 + \cos\theta \cos\theta')
-2\beta(\cos\theta + \cos\theta')\right]
+ {M_Z^2\over 4} \sin\theta \sin\theta' \cos\phi ,\nonumber\\
p_+\cdot k_- = {M_H^2\over 16} \left[
(1+\beta^2)(1 + \cos\theta \cos\theta')
+2\beta(\cos\theta + \cos\theta')\right]
+ {M_Z^2\over 4} \sin\theta \sin\theta' \cos\phi. \nonumber\\
\quad\label{eq:trigo}
\end{eqnarray}
The covariant expression in Eq.(\ref{eq:master}) derived from the usual
Dirac matrix calculation agrees with
the result by the helicity amplitude method in
Eqs.(\ref{eq:MZZ},\ref{eq:rho},\ref{eq:join},\ref{eq:damp}) with
substitutions of Eq.(\ref{eq:trigo}).
The differential CP--odd asymmetry then can be defined as
\begin{equation}
{\cal A}_{u.d.}(E_{p_-}, E_{k_-}, \phi) \equiv
{n(E_{p_-}, E_{k_-}, \phi) - n(E_{p_+}, E_{k_+}, -\phi) \over
n(E_{p_-}, E_{k_-}, \phi) + n(E_{p_+}, E_{k_+}, -\phi)}
={\Delta n(E_{p_-}, E_{k_-}, \phi) \over n_0(E_{p_-}, E_{k_-}, \phi)}\;.
\label{eq:aa}
\end{equation}

To simulate realistic detector acceptance, one has to use a Monte
Carlo calculation based on the differential distribution in
Eq.(\ref{eq:master}).
However, it is instructive to look at the overall asymmetry
integrating over the full ranges of $\cos\theta$ and $\cos\theta'$.
For the CP--odd numerator in ${\cal A}_{u.d.}$
of Eq.(\ref{eq:aa}), we have
\begin{eqnarray}
\Delta n(\phi)\equiv
\int \Delta n d\cos\theta d\cos\theta' =
- {{\cal N}B D M_H^4 \over 2 \cdot 9} &z&(1-4z)^{1\over2}
   \biggl[
           z  \sin2\phi (c_L^2 + c_R^2)^2     \nonumber\\
   &-& {9\pi^2\over 64} (1-2z) \sin\phi   (c_L^2 - c_R^2)^2 \biggr]
\;,
\label{eq:intodd}
\end{eqnarray}
with $z=M_Z^2/M_H^2$.
The first term in the bracket is due to
the interference between $M_{+,+}$
and $M_{-,-}$ and is proportional to $\rho^+_- - \rho^-_+$.
The second term in the bracket is due to
the interference between $M_{0,0}$
and $M_{\pm,\pm}$ and is proportional to
$\rho^0_+ - \rho^0_- + \rho^-_0 - \rho^+_0$.
For the CP--even denominator, we have
\begin{eqnarray}
n_0(\phi)\equiv  \int n_0 \;d\cos\theta d\cos\theta'
&=&   {{\cal N}B^2M_H^4\over 4\cdot 9} \biggl[
      (1- 4 z + 12 z^2 + 2 z^2 \cos2\phi) (c_L^2 + c_R^2)^2 \nonumber\\
&\ &  -{9\pi^2 \over 16} z (1- 2 z)
\cos\phi (c_L^2 - c_R^2)^2 \biggr] \;.
\label{eq:inteven}
\end{eqnarray}
The CP conserving part, $n_0(\phi)$ has been calculated
before\cite{ref:bij,ref:tc}.
Our result agrees with Ref.\cite{ref:tc}.
A sign difference between ours and Ref.\cite{ref:bij}
can be due to different definitions of $\phi$.   
The normalization ${\cal N}$ can be chosen to be
\begin{equation}
{\cal N}=18/[\pi B^2M_H^4(1-4z + 12z^2)(c_L^2+c_R^2)^2]  \;,
\label{eq:norm}
\end{equation}
such that the distribution $n(\phi)=n_0(\phi)+\Delta n(\phi)$
is normalized to one after integration over $\phi$.
Measuring these $\sin\phi$ or $\sin2\phi$ dependences in the
event distribution will establish the CP non--conservation.
To enhance statistics,  we can look at the up-down asymmetry by
integrating again over $\phi$ from 0 to $\pi$,
\begin{equation}
{\cal A}_{u.d.}={ \left(\int_0^\pi-\int_\pi^{2\pi} \right) n(\phi)d\phi
      \over      \int_0^{2\pi}                      n(\phi)d\phi }
={D\over B} {9\pi z(1-4z)^{1\over2}(1-2z) \over 16(1-4z+12z^2) }
            \left( {c_L^2-c_R^2 \over c_L^2+c_R^2} \right)^2 \;.
\label{eq:aintas}
\end{equation}
After the integration, the $\sin2\phi$ term
of the combination $c_L^2+c_R^2=2(c_A^2+c_V^2)$ gives zero while
the surviving $\sin\phi$ term is relatively suppressed by the factor
$c_L^2 - c_R^2 =-4 c_V c_A$ as we anticipated before.
A more realistic integrated asymmetry should
be the one that can preserve the $\sin2\phi$ term.
That can be obtained by
taking the difference in the integration over $[0, \pi/2]$
and the integration over $[\pi/2, \pi]$.
\begin{equation}
{\cal A}_{u.d.}'={ \left(\int_0^{\pi/2}
                 -\int_{\pi/2}^\pi
                 +\int_\pi^{3\pi/2}
                 -\int_{3\pi/2}^{2\pi}
            \right)              n(\phi)d\phi
      \over      \int_0^{2\pi}   n(\phi)d\phi }
=-{D\over B} {4z^2(1-4z)^{1\over2} \over \pi(1-4z+12z^2) }
\;.
\label{eq:aintasy}
\end{equation}
Fig.~2 shows ${\cal A}_{u.d.}$, ${\cal A}_{u.d.}'$ versus $M_H$
per unit of $D/B$.
Following the above formalism, experiments can search for CP
non-conservation, or at least, put a constraint on the parameter $D$.
Note that, in the heavy Higgs limit $z\ll 1$, the $\sin\phi$ contribution
becomes relatively important because of the kinematic suppression factor
$z$ in the other contribution of $\sin2\phi$ in Eq.(\ref{eq:intodd}).
\section{Generalization}
We can generalize our formalism to the case that $l$ and $l'$
are different fermions of different couplings $c_L,c_R$ and $c_L',c_R'$.
The replacements in
Eqs.(\ref{eq:master},\ref{eq:intodd}--\ref{eq:aintas})
are given by the following rules,
\begin{eqnarray}
c_L^4+c_R^4 &\rightarrow&  c_L^2c_L^{'2} +c_R^2c_R^{'2}   \;, \nonumber\\
2c_L^2c_R^2 &\rightarrow&  c_L^2c_R^{'2} +c_R^2c_L^{'2}   \;, \nonumber\\
(c_L^2+c_R^2)^2 &\rightarrow& (c_L^2+c_R^2)({c'}_L^2+{c'}_R^2)\;,
\nonumber\\
  (c_L^2-c_R^2)^2 &\rightarrow& (c_L^2-c_R^2)({c'}_L^2-{c'}_R^2)\;.
\end{eqnarray}
This applies also to the case
$H\rightarrow ZZ \rightarrow l^-l^+ q \bar q$
and $H\to ZZ \rightarrow q \bar q q'\bar q'$.
 
One of the main problems of using the hadronic or semi-hadronic
decay modes of the Higgs boson in searching for CP violating
signals is that experimentally it is impossible to identify charges
of jets originating from hadronization of the partons.  Therefore
one essentially has to smear over the charge information.  Naively one
may think that it will make CP information impossible to
disentangle.  This turns out not the case in the observable 
$O'_{odd}$, which is even under both $\hat C_1$ and $\hat C_2$ 
separately. Thus, in $O'_{odd}$ the contribution of an event and its 
$\hat C_1$--conjugate event will add, while in $O_{odd}$ they will cancel.
Therefore, charge identification is not necessary for the observable
${O'}_{odd}$.   
 
To construct the observable,
we revisit the definition of $\phi$ in the general case.
Each pair of fermions defines a plane.
Given two un--oriented planes one can define the angle
$\phi_{0}$ between planes to be $a$ or $\pi-a$
($ 0 \le a \le \pi$ by definition).
These two choices are not resolved yet.
However, if the common line of the two planes can be physically
assigned a direction associated with one of the planes,
then the two angles $a$ and $\pi-a$ can be distinguished.
For example in the $H \rightarrow ZZ$ decay, see Fig.~1,
we can simply use the vector $\vec k_- + \vec k_+ = \vec k_Z$
to define a direction associated with the $k_-,k_+$ plane.
Then, using the right hand rule, one can rotate along
the axis $\vec k_- +\vec k_+$ and
sweep the plane of $p_-,p_+$ toward the plane of
$k_-,k_+$ and define the resulting angle
$\phi_0\in[0,\pi]$.
It is important to notice that choosing $\vec p_- +\vec p_+$ to define
the angle, instead of $\vec k_- +\vec k_+$, gives the same result.
(Therefore the two fermion pairs can be identical without smearing out
the effect.)
On the other hand, the azimuthal angle $\phi$ defined
in Fig.~1 with full identification varies from 0 to $2\pi$, and the
relation between them is simply $\phi_{0}=$mod$(\phi,\pi)$.
So the event rate at $\phi_{0}=a$ ($0 \le a \le \pi$ by definition)
is the sum of the elementary rates $n(\phi=a) + n(\phi=a+\pi)$,
thus the dependences on $\sin\phi$ in Eq.(\ref{eq:intodd})
and $\cos\phi$ in Eq.(\ref{eq:inteven})  will be washed away.
However, the $\sin2\phi$ term survives to signal the CP violation.
Of course, if one can identify the charges of the final fermions,
as in the case of leptonic modes,
then one can also decode the $\sin\phi$ term as well.
To detect the $\sin2\phi$ dependence, one can use the same
integrated asymmetry, ${\cal A}_{u.d.}'$,
as in Eq.(\ref{eq:aintasy}) except in this case
the sums $\int_0^{\pi/2}+\int_\pi^{3\pi/2}$ and
$\int_{\pi/2}^\pi + \int_{3\pi/2}^{2\pi}$ are automatically taken care of
by the definition of $\phi_{0}$.
Therefore Eq.(\ref{eq:aintasy}) is still
valid even in the case of
$l^-l^+ q \bar q$ or $q \bar q q'\bar q'$ decay modes.
Note that for these modes the asymmetries corresponding to
Eq.(\ref{eq:aintas}) are not doubly suppressed by the small $c_v$ as
in the purely leptonic modes.  Unfortunately, ${\cal A}_{u.d.}$
is not observable due to the lack of hadronic charge identification.
However, one should keep in mind that if one is willing to zoom into the
dependence of the amplitudes on the jet energies
(or, equivalently, on the
angles $\theta$, or $\theta'$) then it may be possible to use energy
identification to replace charge identification to
construct an observable
similar to ${\cal A}_{u.d.}$.
 
Note that for the CP conserving decay, the angular correlation we just
described
can also be used to detect the important $\cos 2\phi$
distribution even for
the semi-hadronic modes,
$l^-l^+ q \bar q$, or the hadronic modes, $q \bar q q'\bar q'$, of the
$H \rightarrow ZZ$ decay.

\section{$H\to W^+W^-$}
Once one understands how the CP--odd correlation
can be decoded in the hadronic
modes of $H \rightarrow ZZ$ decay, it is easy to apply it to
the process $H \rightarrow W^+ W^-$ also.
In this case one can write down a similar
vertex with the $Z$ fields replaced  by the $W$ fields. Since the
coupling of the $W$ boson to a fermion pair is always left handed,
we just set $c_R=c'_R=0$ and $c_L=c_L'=1$. All substitutions are
straightforward. We can simply add the superscript $W$ to
$B$, $D$, $n$, $\Delta n$, ${\cal A}_{u.d.}$ and ${\cal A}_{u.d.}'$
in Section IV to
label the $WW$ mode in the above formulas.
 
In order to determine the decay planes of both $W^\pm$ bosons,
we cannot use the purely leptonic mode $l\bar\nu  \nu \bar l$, where
too much kinematic information is carried away by the missing $\nu$ and
$\bar\nu$. However, we can use the mixed modes $l\bar\nu q\bar q$ or
$\bar l\nu q\bar q$, where kinematics can be fully reconstructed
provided that we have good resolution on jets. Without identifying the
nature of the leading quarks in jets, we are still able to
define uniquely the $\phi_{0}$ as defined in Section V.
This is good enough to measure the $\sin2\phi$
dependence, which is related to ${\cal A'}^W_{u.d}$,
a quantity generalized from
the definition in Eq.(\ref{eq:aintasy}).
In Fig.~3, we show ${\cal A}^W_{u.d.}$,
${\cal A'}^W_{u.d}$ versus $M_H$ per unit of $D^W/B^W$.
Note that, unlike the purely leptonic modes of $H \rightarrow ZZ$ case,
${\cal A}^W_{u.d.}$ is much lareger than ${\cal A'}^W_{u.d}$.
Unfortunately, ${\cal A}^W_{u.d.}$ is not observable
due to lack of charge
identification.  However, as commented earlier for the $ZZ$ case,
it may be possible to replace charge identification
by energy identification
to construct an observable similar to ${\cal A}^W_{u.d.}$.
 
For the purely hadronic mode, there is tremendous background from
non--resonant contributions in the hadronic environment.
The $e^-e^+$ collider may be a cleaner
machine in which the Higgs boson may be produced via the process
$e^- e^+ \to ZH$. Then,  one can use the decay $Z \to l^- l^+$
to tag the recoiling Higgs boson $H$ which turns into a $W^+ W^-$
pair and finally becomes 4 jets, whose CP character
can be decoded using the angular correlation defined earlier.
 
In addition, our study of the angular correlation can be very useful
in investigating the general CP--odd or CP--even correlations in other
reactions, for example, the hadronic final states of the process
$e^+ + e^- \rightarrow W^+ + W^-$\cite{ref:ckp2}.
\section{}
\centerline{$H\to t \bar{t}$}
The possibility of detecting CP violation in $H\to t \bar{t}$ is very
interesting in the sense that the corresponding CP violating couplings
can occur at the tree--level.
Let us look at the phenomenological form of the Yukawa interaction,
\begin{equation}
{\cal L}^{Yukawa}_{CPX}=-(m_t/v) \bar t (A P_L+A ^*P_R) t H
\;.
\label{eq:Yukawa}
\end{equation}
Here $v=(\sqrt{2} G_F)^{-{1\over 2}} \simeq 246 \hbox{ GeV}$.
The complex coefficient $A$ is a combination of model-dependent mixing
angles.  The CP violating effect is
proportional\cite{ref:Weinberg,ref:Barr}
to  Im~$A^2 = 2 \hbox{Im } A \hbox{Re }A$.
%
In Higgs boson production at a high energy collider, one expects the
CP$\hat{T}$-even, CP violating effect to be observable in the angular
asymmetry as before.  It can in principle be tree--level effect and
therefore can be very significant.
However, in reality, its signal is harder to decode. The $t \bar{t}$
pair decay into $W^+ b W^- \bar{b}$.
If both $W$ bosons decay leptonically,
one faces the problem of identifying the Higgs boson event
with the missing neutrinos. If a $W$ boson decays hadronically one
has to deal with its multi-jet final state.
Assuming that one can identify the jets associated the $W$'s
then one can use the momenta of $W$'s and $b, \bar{b}$ to define the
decay plane of $t \bar{t}$.  Then the C--even angular correlation
discussed in previous section can be used to decode CP violation. In
hadronic colliders this is probably too hard to achieve.  In a leptonic
machine this may be possible only if the Higgs is produced in the
futuristic NLC (such as EE500)
or $\gamma\gamma$ colliders\cite{ref:NLC}.
 
In this section, we will give the full differential form of the decay
probability, which is useful for future experimental simulation.
Without loss of generality, we look at the process $H\to t\bar t$, with
$t\to b \bar e \nu$ and $\bar t\to \bar b e\bar\nu$. Formulas for other
processes can be obtained by simple substitutions. Our result is based
on the standard V$-$A coupling for the $t$ decay .
\begin{eqnarray}
\sum_{spin}\vert M \vert^2&=&
8g^{8}m_t^2 b\cdot\nu \bar b\cdot \bar \nu
\Biggl\{  (2e\cdot H \bar e\cdot H-e\cdot\bar e H^2) A_I^2 \nonumber\\
&+&
[2e\cdot (t-\bar t) \bar e\cdot (t-\bar t)
-e\cdot\bar e (t-\bar t)^2]A_R^2
\nonumber\\
&+&4\epsilon(e,\bar e,t,\bar t)A_IA_R \Biggr\}
(m_t/v)^2
\vert {\cal P}_t
      {\cal P}_{\bar t}
{\cal P}_{W^+}
{\cal P}_{W^-} \vert^2   \;.
\label{eq:Htt}
\end{eqnarray}
For brevity, it is understood that we use the particle symbol to denote
its corresponding momentum. The last four propagators, ${\cal P}$'s,
can be treated quite easily in the narrow--width approximation;
simply  replace the virtual momenta by their on-shell values, and
the integrations over the squares of virtual momenta will give an
overall factor
$$\left({\pi\over m_t\Gamma_t} {\pi\over M_W\Gamma_W} \right)^2   \;.$$
Inspecting Eq.(\ref{eq:Htt}), we know that $\epsilon(e,\bar e, t,\bar t)$
is an appropriate CP--odd observable. Since
$$\epsilon(e,\bar e, t,\bar t)
=-\case 1/2M_H \vec p_e\times\vec p_{\bar e}
       \cdot (\vec p_t   -  \vec p_{\bar t}) \;, $$
in the $H$ rest frame,
CP information resides in the relative azimuthal angle between
$\vec p_e$ and $\vec p_{\bar e}$ with respect to the axis
$\vec p_t   -  \vec p_{\bar t}$.
In contrast to Eq.(\ref{eq:intodd}) there is only one CP--odd
angular distribution in Eq.(\ref{eq:Htt}).  This is because the top quark
is a spin--$\case1/2$ object and has only two helicity states.
The corresponding CP--odd variable is thus
proportional only to $\rho^+_- - \rho^-_+$ where $\pm$ represents the
sign of the top helicities.
 
In the future it may be possible to detect other similar CP--odd
correlations in process such as
$\gamma + \gamma \rightarrow t + \bar{t} +H$.
\section{}
\centerline{$e^+ e^- \rightarrow Z + H$}
The analysis given above can be easily applied to the
equally important process of the Higgs production as well.
Here we shall simply investigate a typical one which is promising.
The Higgs boson may be produced\cite{ref:Ma}
through the $Z$ bremsstrahlung in
future high energy $e^-e^+$ colliders and
the CP information can be carried
over by the lepton pair from the $Z$ decay.
To visualize the process of $e^-(p_-)e^+(p_+) \to Z(P)H$ where
$Z\to l(k_-)\bar l(k_+)$, we sketch the reaction configuration
in  Fig.~4. The final on--shell $Z$ boson is produced at a scattering
angle $\Theta$.  Its subsequent decay into a lepton pair is described by
the polar angle $\theta$ and the azimuthal angle $\phi$ of $l$ in the
$Z$ rest frame with its $\hat z$ axis opposite to
$\vec P =\vec k_- +\vec k_+$. We
focus our study in the case that the $Z$ boson, produced simultaneously
with the Higgs boson, is on--shell and we
will use the narrow--width  approximation to handle its decay process.
 
The virtual gauge boson in the $s$--channel can be $\gamma^*$ or $Z^*$.
It has the momentum $P^*=p_-+p_+$.
First, we write down the relevant form factors
\begin{equation}
  iM_{Z^*(P^*,\eta_*)\to Z(P,\eta)H}
= ig_H[B\eta \cdot \eta_* +
D^{Z^*} M_H^{-2}\epsilon(\eta,\eta_*,P,-P^*)]    \;,
\end{equation}
and
\begin{equation}
  iM_{\gamma^*(P^*,\eta_*)\to Z(P,\eta)H}
= ig_H[ D^{\gamma^*} M_H^{-2}\epsilon(\eta,\eta_*,P,-P^*)]         \;.
\end{equation}
The form factor $D^{Z^*}$ is related to $D$ in Section II when one
of the $Z$'s becomes virtual. We put in a negative sign in front of
$P^*$ so that the momentum flow is consistent with that defined in
Eq.(2). The CP--odd term of $D^{\gamma^*}$ provides
another source of CP violation. Other higher--dimensional CP--even terms
are omitted  in our discussion.
The overall transition probability is
\begin{eqnarray}
\sum_{spin} |M|^2 &=&\left({4 e^2 g_H\over s-M_Z^2}\right)^2
\left| {1\over P^2-M_Z^2+i\Gamma_ZM_Z} \right|^2
\nonumber\\
  &\times& \Biggl\{
    B^2 \left[ (c_L^4+c_R^4)  p_-\cdot k_-   p_+\cdot k_+
  +2 c_L^2 c_R^2  p_-\cdot k_+   p_+\cdot k_- \right]
\nonumber\\
 &+& B\epsilon(p_-,p_+,k_-,k_+)\biggl[
    (c_L^2 \bar c_L^2 + c_R^2 \bar c_R^2) (p_-\cdot k_- + p_+\cdot k_+)
\nonumber\\
 &-&  (c_L^2 \bar c_R^2 + c_R^2 \bar c_L^2) (p_-\cdot k_+ + p_+\cdot k_-)
   \biggr]/M_H^2 \Biggr\}  \;,
\label{eq:ZH}
\end{eqnarray}
where the $D$ form factors has been lumped into the following modified
couplings,
\begin{eqnarray}
\bar c_L^2 &=& c_L^2 D^{Z^*}-c_L(1-M_Z^2/s)D^{\gamma^*}  \;,\nonumber\\
\bar c_R^2 &=& c_R^2 D^{Z^*}-c_R(1-M_Z^2/s)D^{\gamma^*}  \;.
\end{eqnarray}
The $Z$ couplings for the electron or the muon are
$c_L=(-\case1/2+\sin^2\theta_W)/(\sin\theta_W\cos\theta_W)$ and
$c_R=\tan\theta_W$.
The differential cross section is
\begin{equation}
{d\sigma\over d\cos\Theta d\cos\theta d\phi}=
{\int \sum |M|^2 \lambda^{1\over 2}dP^2 \over 8192\pi^4 s} \;.
\end{equation}
The kinematic factor $\lambda^{\case1/2}=2|\vec p_H|/\sqrt{s}$
is defined in the $e^-e^+$ CM frame.
The integration over $P^2$ can be simplified in the narrow width
approximation. One just replaces $P^2$ by its on--shell value $M_Z^2$ and
substitutes the corresponding propagator squared by $\pi/(M_Z\Gamma_Z)$.
Under CP transformation, the configuration transforms as follows,
\begin{equation}
(\cos\Theta,\cos\theta,\phi)
\to (\cos(\pi-\Theta),\cos(\pi-\theta),-\phi) \;.
\end{equation}
If the differential cross section does not respect the symmetry of this
transformation, CP is violated. The presence of the $\epsilon$ term in
Eq.(\ref{eq:ZH}) will produce such CP violation.
One can integrate Eq.(\ref{eq:ZH}) over the polar angle $\theta$ of $l$,
and obtain a $\phi$ distribution,
\begin{equation}
{d\sigma\over d\cos\Theta d\phi} =
{\alpha^2\over64s}
\left({\alpha M_Z (c_L^2+c_R^2)
\over 6\Gamma_Z}\right)
{(c_L^2+c_R^2)B^2 \over \sin^2\theta_W\cos^2\theta_W}
{\Sigma_0+\Delta\Sigma \over (1-M_Z^2/s)^2}
\lambda^{1\over2}
\;.
\end{equation}
The second factor involving the $Z$ width is just the branching fraction
BF$(Z\to l\bar l)$.
We purposely separate the CP--even part $\Sigma_0$ and
the CP--odd part $\Delta\Sigma$ in the above formula. The CP--even
part is
\begin{eqnarray}
     \Sigma_0&=&\lambda\sin^2\Theta +
                   {2M_Z^2\over s}(4+\sin^2\Theta \cos2\phi)
             \nonumber\\
              &+& {3\pi M_Z\over 2\sqrt{s}}
                  (1+{M_Z^2\over s}-{M_H^2\over s})
                  \sin\Theta\cos\phi {(c_L^2-c_R^2)^2
                                 \over(c_L^2+c_R^2)^2}
\;.
\end{eqnarray}
The CP--odd part consists of two terms,
\begin{eqnarray}
\Delta\Sigma  = (a_2 \sin^2\Theta
                 \sin2\phi
               +a_1 \sin\Theta\sin\phi)\lambda^{1\over2}
\;.
\end{eqnarray}
Both terms of $\sin\phi$ and $\sin2\phi$ are
CP violating. Their coefficients are
\begin{equation}
a_1={(c_L^2-c_R^2)(\bar c_L^2-\bar c_R^2)
                  \over B(c_L^2+c_R^2)^2}
                 {3\pi M_Z\sqrt{s}\over 4M_H^2}
                \left(1+{M_Z^2\over s}-{M_H^2\over s}\right),\
a_2={2\over B}{M_Z^2   \over M_H^2}
{\bar c_L^2+\bar c_R^2\over c_L^2+c_R^2}
\;.
\end{equation}
Note that $\lambda = (1+{M_Z^2/s}-{M_H^2/s})^2 - 4{M_Z^2/s}$.
We can define the integrated asymmetries as before,
\begin{equation}
{\cal A}^{ZH}_{u.d.} \equiv {1\over \sigma} \int_{-1}^1d\cos\Theta
            \left(\int_0^\pi
                 -\int_\pi^{2\pi}
            \right) d\phi {d\sigma\over d\cos\Theta d\phi}
={3 a_1 \lambda^{1\over2} \over 4 (\lambda+12M_Z^2/s)}
\;,
\end{equation}
\begin{eqnarray}
{{\cal A}'}^{ZH}_{u.d.} &\equiv& {1\over\sigma}
\int_{-1}^{1}d\cos\Theta
            \left(\int_0^{\pi/2}
                 -\int_{\pi/2}^\pi
                 +\int_\pi^{3\pi/2}
                 -\int_{3\pi/2}^{2\pi}
            \right)d\phi {d\sigma\over d\cos\Theta d\phi}
                                                 \nonumber\\
&=&{2 a_2 \lambda^{1\over2} \over \pi(\lambda+12M_Z^2/s)}
\;.
\end{eqnarray}
In Fig.~5, we plot
${\cal A}_{u.d}^{ZH}$ and ${{\cal A}'}_{u.d}^{ZH}$
per unit of $D^{Z^*}/B$ or $D^{\gamma^*}/B$ versus $\sqrt{s}$ for
the case $M_H=80$ GeV.
Both the CP--even and the CP--odd contributions to the cross section
decrease with increasing $s$ when $s$ is far above the threshold. However
the CP--even part decreases slightly faster and therefore both
${\cal A}_{u.d}^{ZH}$ and ${{\cal A}'}_{u.d}^{ZH}$ are increasing
functions of $s$.  In addition,
${\cal A}_{u.d}^{ZH}$ increases faster than ${{\cal A}'}_{u.d}^{ZH}$
because of the factor $\sqrt{s}$ in $a_1$.
The process rate is proportional to $s^{-1}$ for large $s$, but since
every reaction, including the background,
is proportional to $s^{-1}$, we conclude that
higher energy is preferable to detect CP violation
provided one can get high enough luminosity.

An analysis similar to what we have done here can also be applied to
processes, like
$e^-e^+ \rightarrow e^- e^+ +(\gamma^*, Z^*) +(\gamma^*, Z^*)
        \rightarrow  e^- e^+  H$.
Of course even in $e^-e^+ \rightarrow Z H
                          \rightarrow l \bar{l} b \bar{b}$,
one can also analyze CP violation
in the Higgs decay by zooming into the angular correlations of
decay products from the final $b$ quark jets.
\section{Higgs Model}
In renormalizable models the CP--odd couplings for both the $H\to ZZ$
and the $H\to W^+ W^-$ modes can be induced only at the loop level
because they are higher dimensional operators. As the $\epsilon$ symbol
in Eq.(\ref{eq:MZZ}) only occurs through  a fermion loop
in perturbative calculations, it is natural to use the top quark as the
internal fermion for its potentially  large Yukawa coupling, see
Eq.(\ref{eq:Yukawa}).
 
In this section, we will show the
CP violating form factors $D$ and $D^W$
in processes $H\to ZZ$ and $H\to WW$
based on the CP non--conserving Yukawa
interaction in Eq.(\ref{eq:Yukawa}).
The one--loop diagram of interest is shown in Fig.~6.
We use the method of dispersion relations to find the form factor
$D$ for the process $H\to ZZ$.  The virtual mass $s$ of the Higgs boson
is analytically extended beyond the on--shell value $M_H^2$. In
Ref.\cite{ref:ck1}, we have the absorptive part,
\begin{equation}
\hbox{Im}\ D(s)
= -{3\sqrt{2} \over 4\pi }G_F m_t^2 \hbox{Im\ }A
    {\beta_t\over \beta_Z^2}
    \Bigl[1+{K\over 4\beta_t\beta_Z}-
        (1-{8\over3}\sin^2\theta_W)^2 {K\beta_Z\over 4\beta_t} \Bigr]
{M_H^2\over s}
       \;,
\end{equation}
with $\beta_t^2=1-4m_t^2/s$, $\beta_Z^2=1-4M_Z^2/s$,
and the logarithmic factor
\begin{equation} K=\log\left\vert {1+\beta_Z^2-2\beta_t\beta_Z
                        \over 1+\beta_Z^2+2\beta_t\beta_Z}
     \right\vert  \;.
\end{equation}
The dispersive part is obtained by the Kramers--Kronig relation.
\begin{equation}
\hbox{Re }D(s)={1\over\pi}P\int^\infty_{4M_Z^2}
                  {\hbox{Im }D(s') \over s'-s} ds' \;.
\end{equation}
The symbol $P$ denotes the principal value of the singular integral.
Fig.~7 shows typical sizes of Re~$D$ for various cases.
It is of order of $10^{-2}$ to $10^{-3}$ in general.
We must keep in mind that we only show the Higgs boson contribution
in the perturbative regime. In scenarios where the Higgs bosons
interact strongly, the CP violating effect in the colliders could
be much larger.
 
Similarly, the form factor $D^W$ for the process $H\to W^+W^-$
can be obtained with the following absorptive amplitude in
Ref.\cite{ref:ck1}.
\begin{equation}
\hbox{Im}\ D^W(s)
   = -{3\sqrt{2}   \over 4\pi} G_F m_t^2 \hbox{Im}\ A
          {\beta_t\over \beta_W^2}
    L(\beta_t,\beta_W)
  {M_H^2\over s}
       \;,
\end{equation}
with $\beta_W^2=1-4M_W^2/s$ and the corresponding logarithmic expression,
\begin{equation}
         L(x,y) \equiv
           1+{x^2-y^2\over 2xy}
       \log \left\vert {x-y \over x+y} \right\vert \;.
\end{equation}
Fig.~8 shows typical sizes of Re~$D^W$ for various cases.
\section{Conclusion}
 
We have shown that the angular asymmetry of the
prompt lepton events originating in Higgs decay
are sensitive to CP violation.
Useful CP--odd variables are constructed in probing
the dispersive parts of the CP violating form factors.
We have demonstrated that the charge identification of jets
is not necessary in order to look for CP signals in $H\to ZZ$ or
$H\to W^+W^-$.
The expected event distributions which we give in their full differential
forms are useful for future experimental simulation. Most of
our approach to the problem is model independent, and so it is
suitable for experimental search for the CP--odd form factors.
 
While we were finalizing this manuscript,
we came across two recent preprints
on the subject of CP violation in the
Higgs decay\cite{ref:SoniXu,ref:HeMa}.
Both papers have some overlaps with our work.   We have noticed that
our conclusion regarding the effect of having the identical particles
($H\to ZZ\to l\bar l l \bar l$) differs from Ref.\cite{ref:SoniXu}.
Also, some sign differences between 
our Eqs.(\ref{eq:intodd},\ref{eq:inteven})
and the corresponding ones in \cite{ref:SoniXu}
can be attributed to the difference in definitions of $\phi$.   
 
This work is supported in parts by the U.S. Department of Energy.
DC wishes to thank L. Wolfenstein for many fruitful discussions.

\eject
%
\section*{Figure Captions}
\begin{itemize}
\vskip .12cm
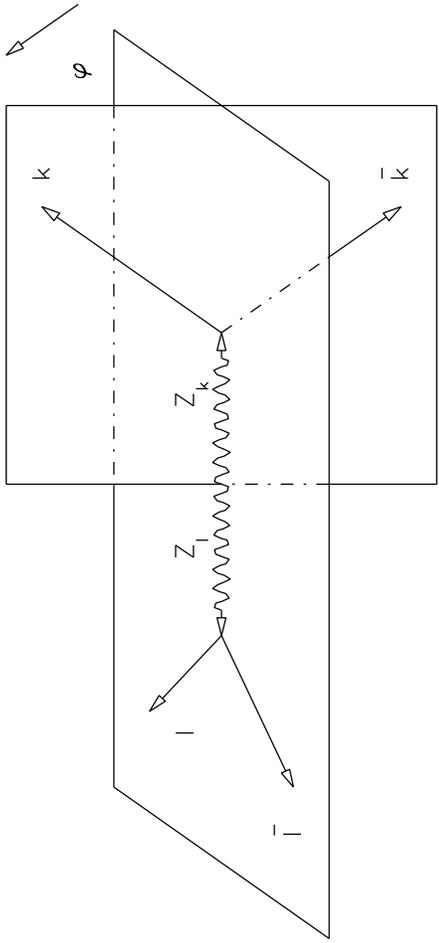
\item[Fig.1] A typical decay configuration,
$H\to ZZ \to l(k_-) \bar l(k_+) l'(p_-) \bar l'(p_+)$.
\vskip .12cm
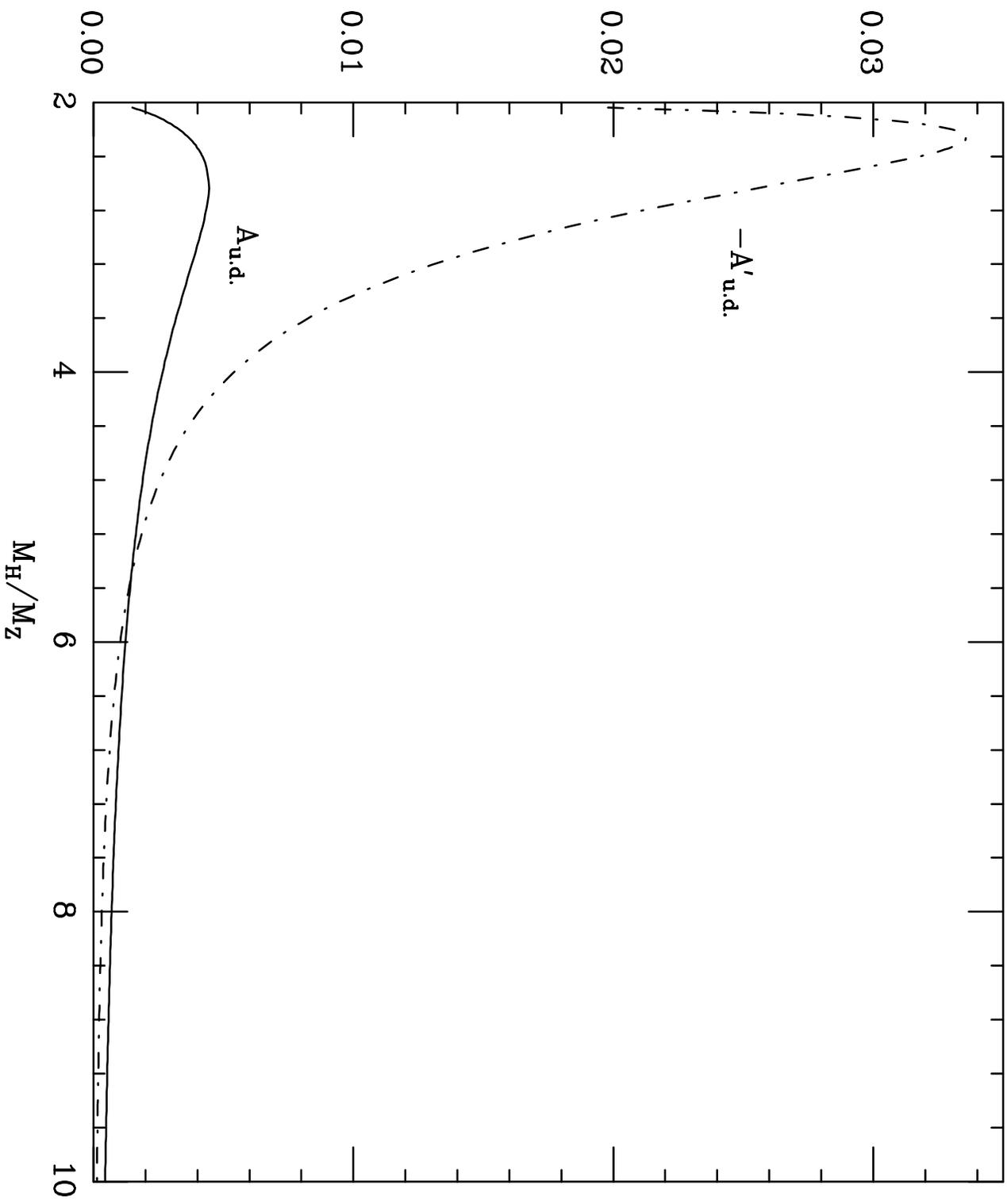
\item[Fig.2] ${\cal A}_{u.d.}$, ${\cal A}_{u.d.}'$ versus $M_H/M_Z$
per unit of $D/B$.
\vskip .12cm
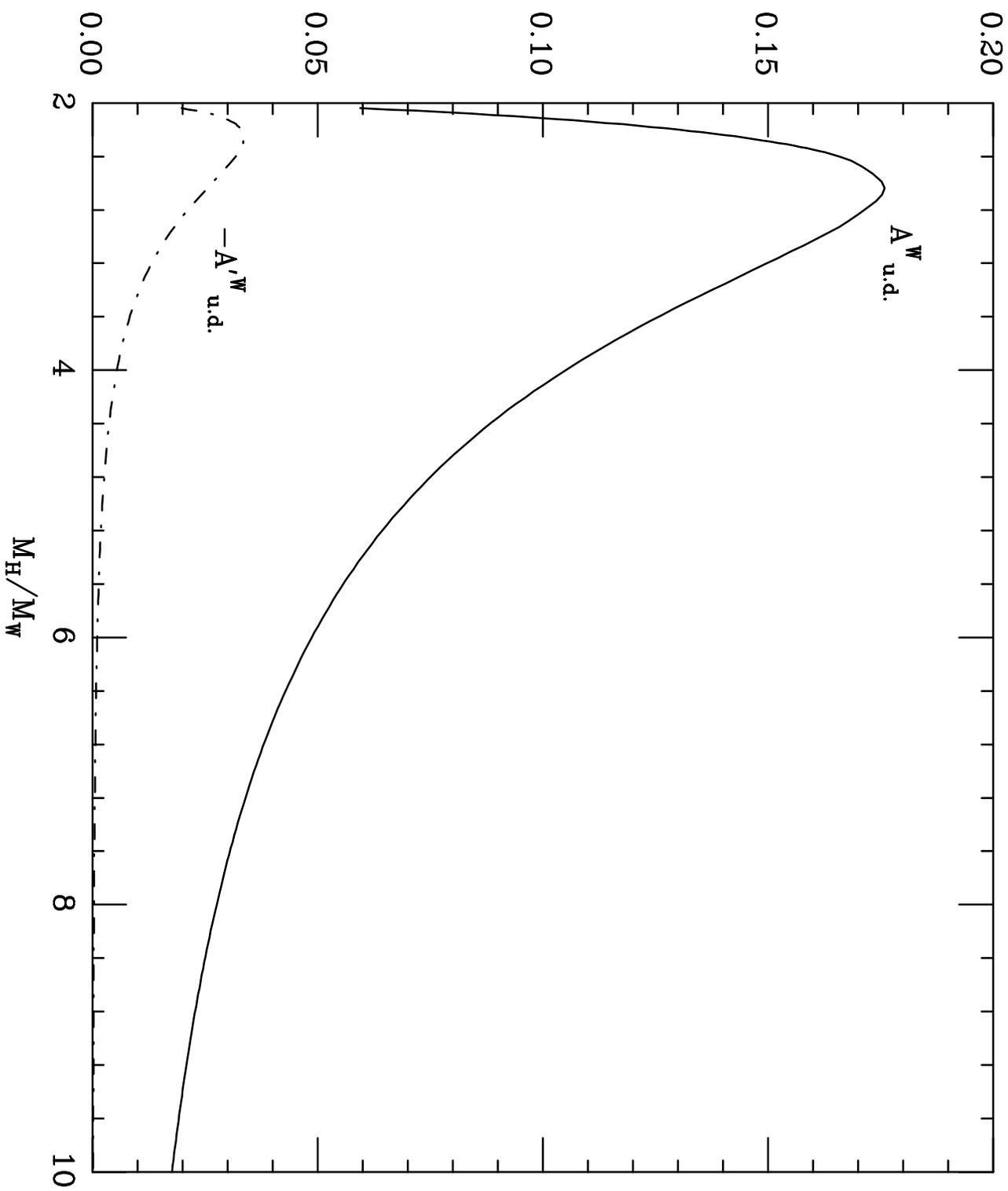
\item[Fig.3] ${\cal A}^W_{u.d.}$,
${\cal A'}^W_{u.d}$ versus $M_H/M_Z$ per unit of $D^W/B^W$.
\vskip .12cm
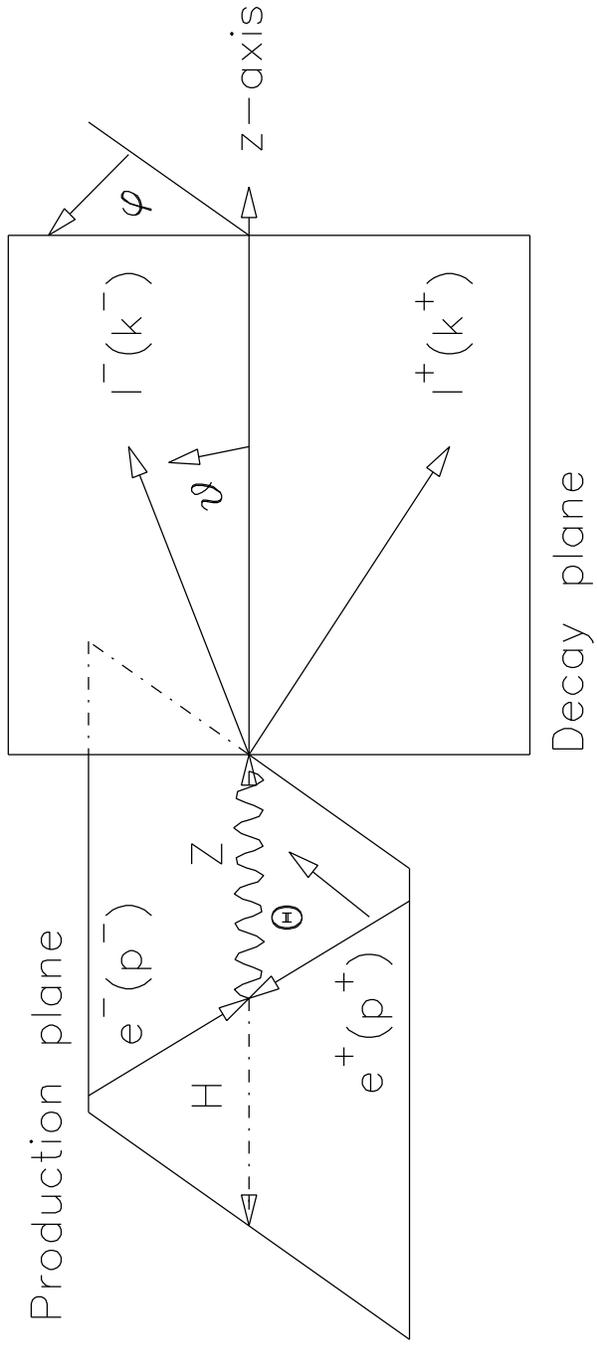
\item[Fig.4] The configuration of $e^-(p_-)e^+(p_+) \to Z(P)H$
where $Z\to l(k_-)\bar l(k_+)$. Note that $\vartheta$ is defined
in the rest frame of $Z$ and is only represented schematically here.
\vskip .12cm
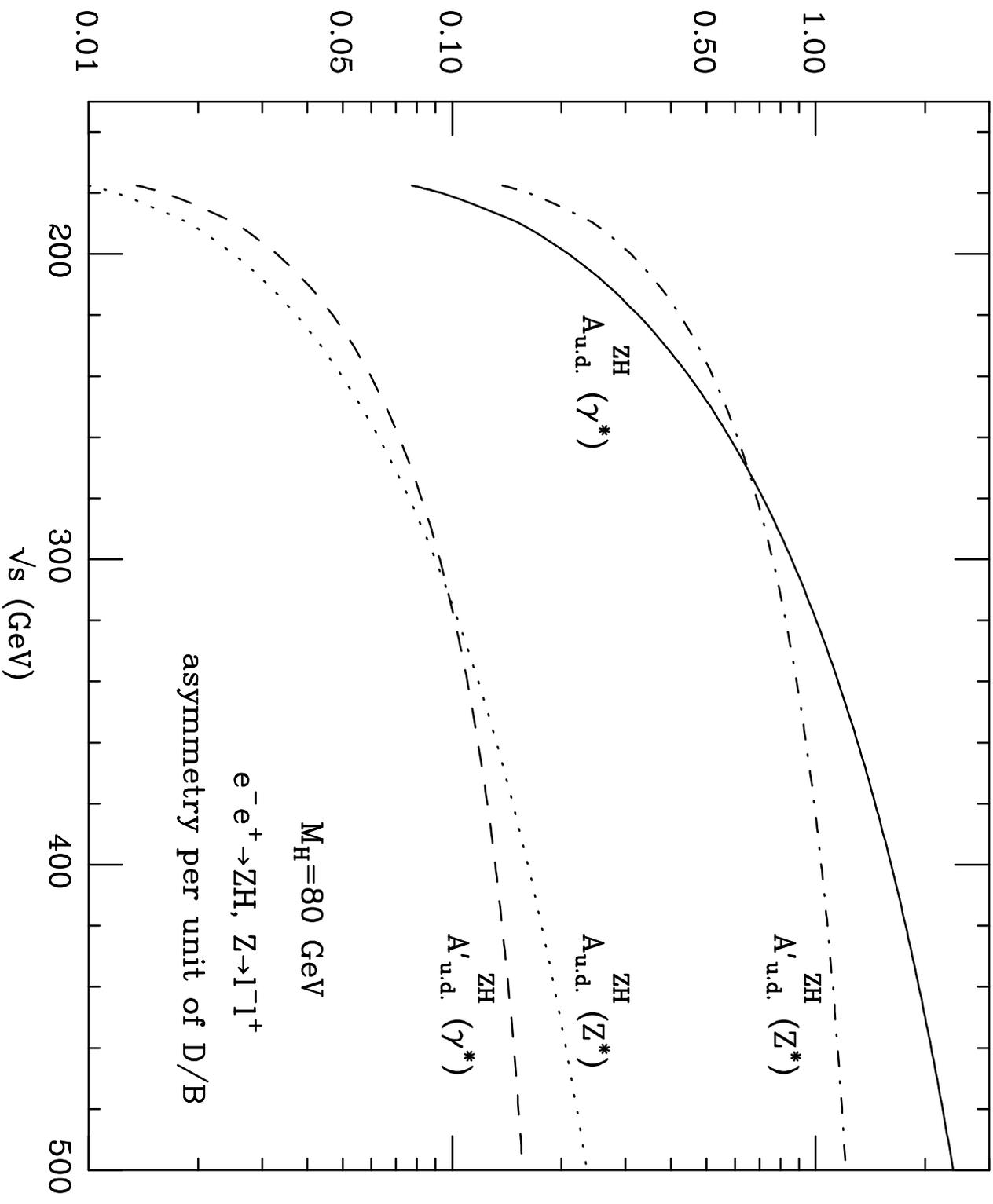
\item[Fig.5] 
${\cal A}_{u.d}^{ZH}$ and ${{\cal A}'}_{u.d}^{ZH}$
per unit of $D^{Z^*}/B$ or $D^{\gamma^*}/B$ versus
$\sqrt{s}$ for $M_H=80$ GeV.
\vskip .12cm
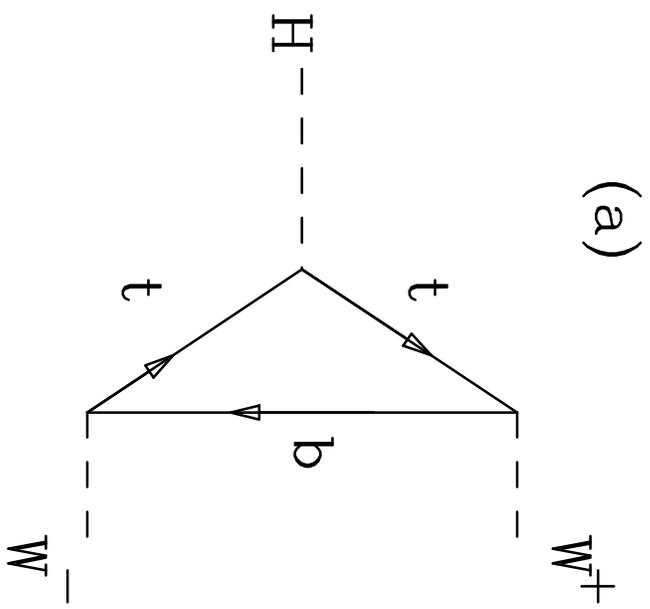
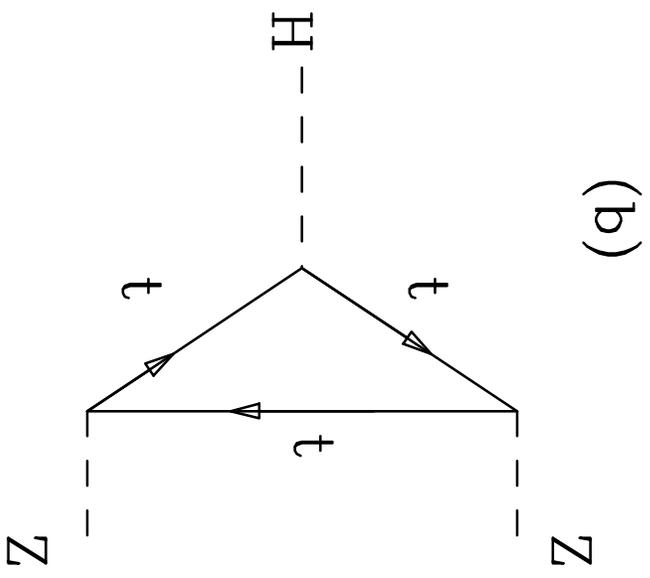
\item[Fig.6] The one--loop diagrams which give the CP violating form factors
$D$ via the intermediate $t$ quark.
\vskip .12cm
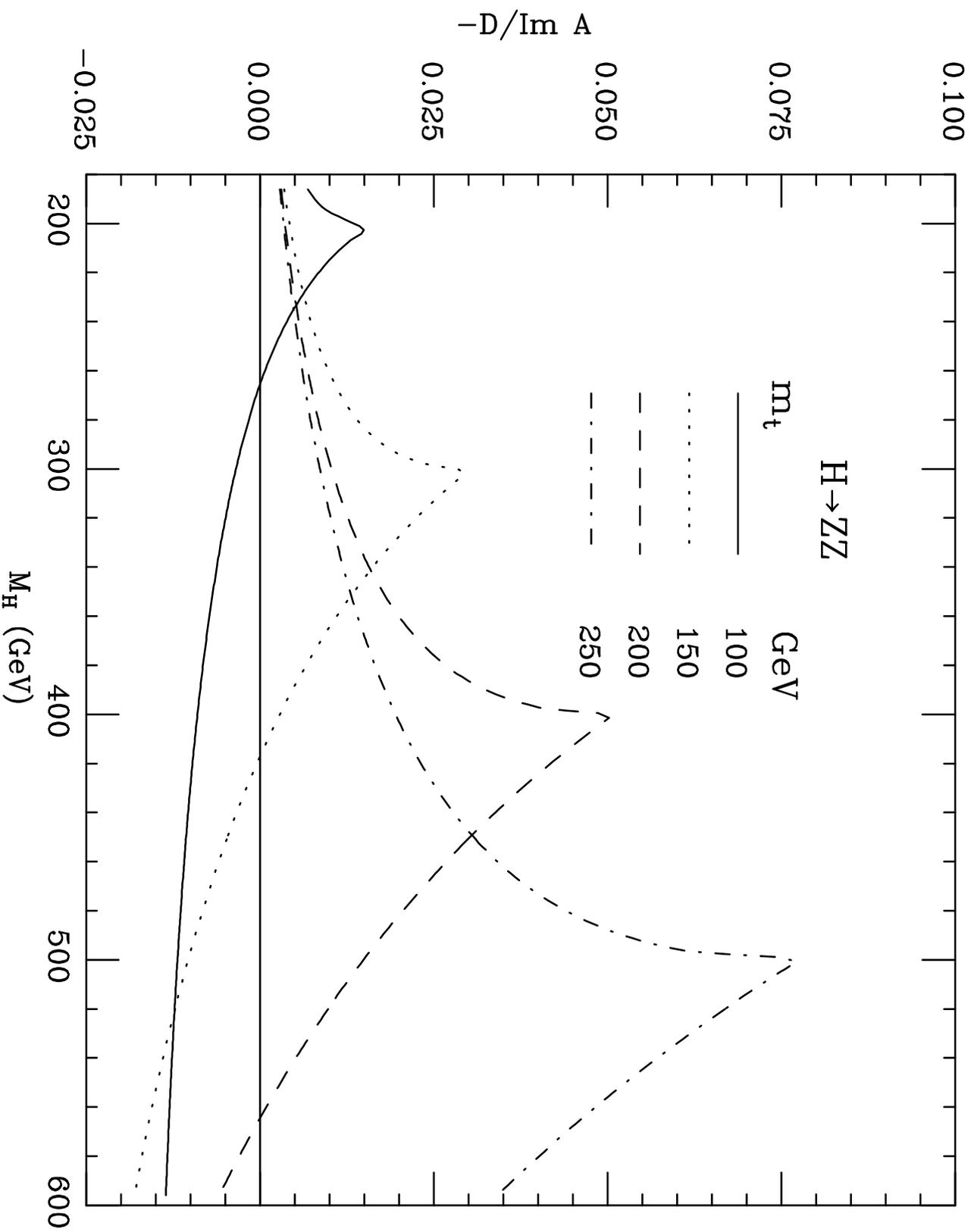
\item[Fig.7] Typical sizes of Re~$D$ for various cases.
\vskip .12cm
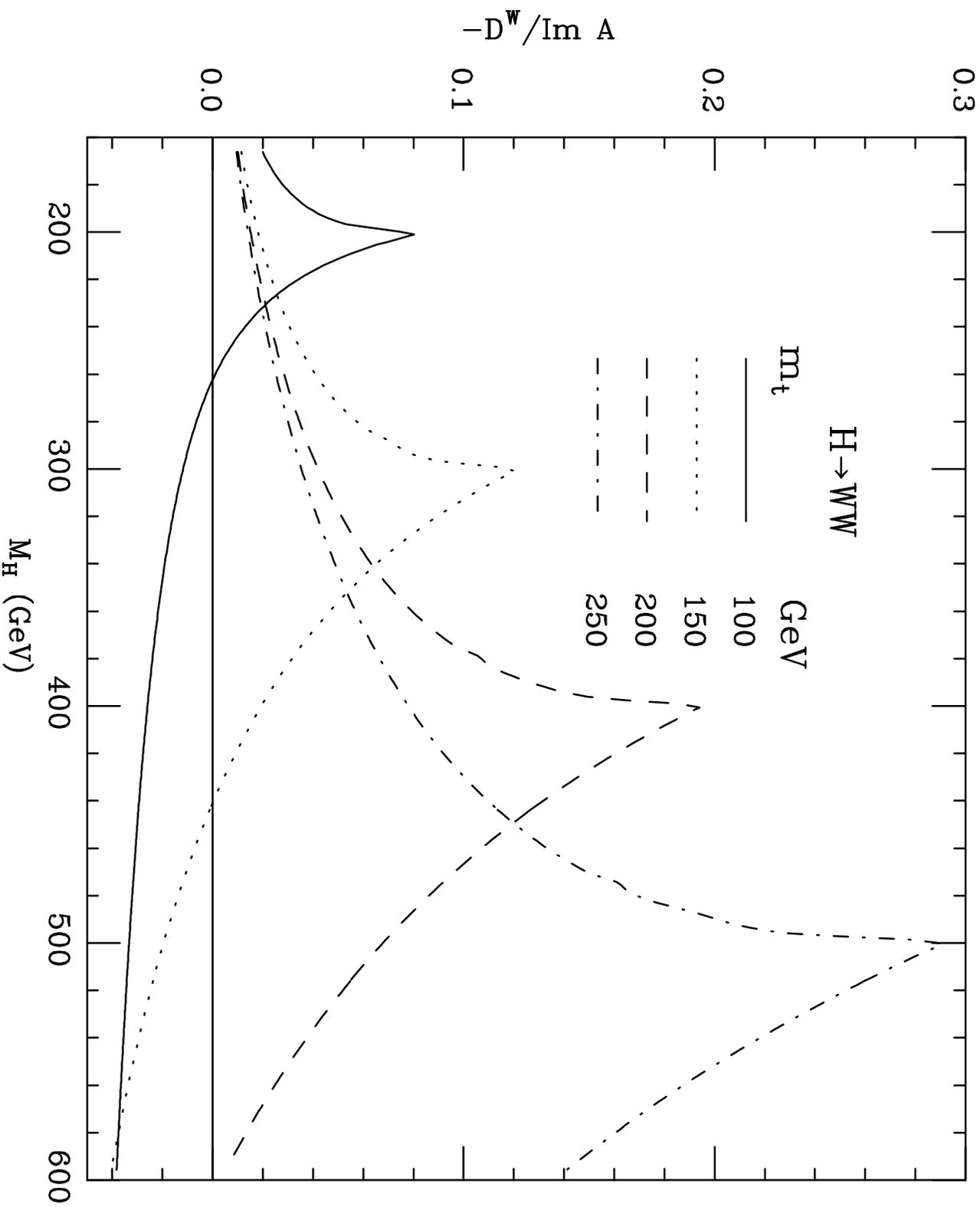
\item[Fig.8] Typical sizes of Re~$D^W$  for various cases.
\end{itemize}
\end{document}